\documentclass[aps,prl,reprint,superscriptaddress]{revtex4-2}

\usepackage[english]{babel}
\usepackage{ulem}
\usepackage[version=3]{mhchem} 
\usepackage{graphicx}
\graphicspath{{msfigure/}} 
\usepackage{amsmath,amssymb}
\usepackage{bm}
\usepackage{siunitx}
\DeclareSIUnit{\rad}{rad}
\DeclareSIUnit{\oe}{Oe}
\usepackage{booktabs}
\usepackage{array}
\usepackage{cancel}
\usepackage{xcolor}
\usepackage{hyperref}
\hypersetup{
    colorlinks=true,
    linkcolor=blue,   
    citecolor=blue,  
    urlcolor=cyan,     
    linktoc=all        
}

\begin{document}

\newcommand{\thetitle}{Nonplanar qubit with tunable gauge symmetry}
\title[\thetitle]
  {\thetitle}
\author{Muqing Yu}
\affiliation{James Franck Institute, University of Chicago, Chicago, Illinois 60637, USA}
\affiliation{Department of Physics, University of Chicago, Chicago, Illinois 60637, USA}

\author{Han Bi}
\affiliation{Department of Physics and Astronomy, Purdue University, West Lafayette, Indiana, 47907, USA}

\author{Hengli Lo}
\affiliation{James Franck Institute, University of Chicago, Chicago, Illinois 60637, USA}
\affiliation{Department of Physics, University of Chicago, Chicago, Illinois 60637, USA}

\author{Vishvesha Sridhar}
\affiliation{James Franck Institute, University of Chicago, Chicago, Illinois 60637, USA}
\affiliation{Department of Physics, University of Chicago, Chicago, Illinois 60637, USA}

\author{Guilherme Delfino}
\affiliation{Department of Physics and Astronomy, Purdue University, West Lafayette, Indiana, 47907, USA}

\author{Dmitry Green}
\affiliation{Department of Physics, Boston University, Boston, Massachusetts 02215, USA}
\affiliation
{AppliedTQC, New York, NY 10065, USA}

\author{Claudio Chamon}
\affiliation{Department of Physics and Astronomy, Purdue University, West Lafayette, Indiana, 47907, USA}
\affiliation{Purdue Quantum Science and Engineering Institute,
Purdue University, West Lafayette, Indiana, 47907, USA}

\author{Nadya Mason}
\affiliation{Pritzker School of Molecular Engineering, University of Chicago, Chicago, Illinois 60637, United States}
\affiliation{Department of Physics, University of Chicago, Chicago, Illinois 60637, USA}

\author{Andrew P. Higginbotham}
\email{ahigginbotham@uchicago.edu}
\affiliation{James Franck Institute, University of Chicago, Chicago, Illinois 60637, USA}
\affiliation{Department of Physics, University of Chicago, Chicago, Illinois 60637, USA}


\begin{abstract}
Circuit quantum electrodynamics embeds Josephson junction qubits within superconducting cavities \cite{Blais2021-hi}, and has emerged as a leading approach to quantum computing \cite{Krantz2019-sn,Kjaergaard2020-rs} and quantum simulation \cite{Houck2012-kq,Fitzpatrick2017-aa,Kollar2019-qu,Ma2019-aa}.
Despite the many permutations of circuit geometry that have been explored~\cite{Schreier2008-aa,Houck2008-aa,Manucharyan2009-xc,Larsen2015-aa,Janvier2015-aa,Gyenis2021-aa}, Josephson connectivities have so far been planar, making them effectively low-dimensional.
Here we show that a non-planar qubit -- a $3\times3$ crossbar Josephson array -- gives rise to flux-tunable $\mathbb{Z}_3$ combinatorial gauge symmetry (CGS), potentially enabling spin-liquid behavior when networked into a lattice \cite{Chamon2020-oz,Yang2021-uj}.
The observed excitation spectrum shows excellent agreement with predictions from a neural network trained to generate variational quantum states \cite{Pfau2024NES}, demonstrating that we have predictive power over our high-dimensional quantum system.
Fine-structure splittings near the CGS point are compatible with weak tunneling or symmetry breaking due to experimental imperfections.
We additionally use the superconducting cavity to externally induce symmetry breaking, observing a restoration of symmetry at the CGS point where ground states differ only by a $\mathbb{Z}_3$ phase.
This work initiates a general program exploring lattice gauge theories \cite{Green2023SciPost, Yu2024SciPost, Yu2025PrB} using the toolbox of circuit quantum electrodynamics.
More broadly, introducing non-planar Josephson connectivities opens a vast space for experimental and theoretical exploration of structures in almost any imaginable dimensionality and geometry \cite{Dede2026-gb}.
\end{abstract}

\maketitle

\section{Introduction}
Planar arrays of Josephson junctions have long been used for experimental studies of the two-dimensional \cite{Geerligs1989-cy,Eley2012-aa,Bottcher2018-zl,Leonard2024-mr} and one-dimensional \cite{Chow1998-aa,Kuzmin2019-aa,Mukhopadhyay2023-aa} superconductor-insulator phase transitions, dissipative phase transitions \cite{Rimberg1997-aa}, and beyond \cite{Sohn1993-aa,Shea1997-xz,Bondar2025-dl}. 
However, existing planar geometries place strong constraints on connectivity: each superconducting node can couple to a limited number of neighbors, constraining the range and topology of the interactions that can be engineered.

Crossbar Josephson arrays, in which two orthogonal sets of wires intersect with a junction at every crossing, lift this restriction.
The geometry naturally realizes all‑to‑all bipartite coupling between the two sets of wires, producing long‑range interactions, frustration, and a degree of connectivity control that is impossible with planar layouts \cite{Dede2026-gb}.
A particularly elegant illustration is the $3\times3$ ``waffle'', where three horizontal wires (matter wires) and three vertical wires (gauge wires) form 9 junctions, realizing the non-planar graph $K_{3,3}$.
When a critical magnetic flux of $\Phi/\Phi_0=1/3$ threads each elementary loop, the waffle acquires an exact $\mathbb{Z}_3$ combinatorial gauge symmetry (CGS) \cite{Chamon2020-oz,Yang2021-uj}, and its Josephson potential possesses 6 degenerate wells.
Ground states in each well are eigenstates of a star operator (the product of three $\mathbb{Z}_3$ clock operators), which is an elementary local building block of spin-liquid states~\cite{Kitaev2003fault}.
When assembled into an appropriately connected lattice, these $3 \times 3$ waffles can therefore realize long-range entangled phases of matter~\cite{Yang2021-uj}.

Before scaling to a whole lattice, the essential microscopic ingredient -- a single waffle with $\mathbb{Z}_3$ CGS -- must be validated.
Similar crossbar arrays were studied earlier using transport measurements, with excellent understanding of the transition temperature \cite{Sohn1993-aa,Shea1997-xz}.
However, more subtle effects associated with degenerate minima -- of key interest here -- were invisible due to an inherent limitation of the transport probes available at the time: the phases unlock as soon as a small current is applied, and individual metastable states cannot be resolved \cite{Shea1997-xz}.

\begin{figure*}[!t]
    \centering
    \includegraphics[width=1\linewidth]{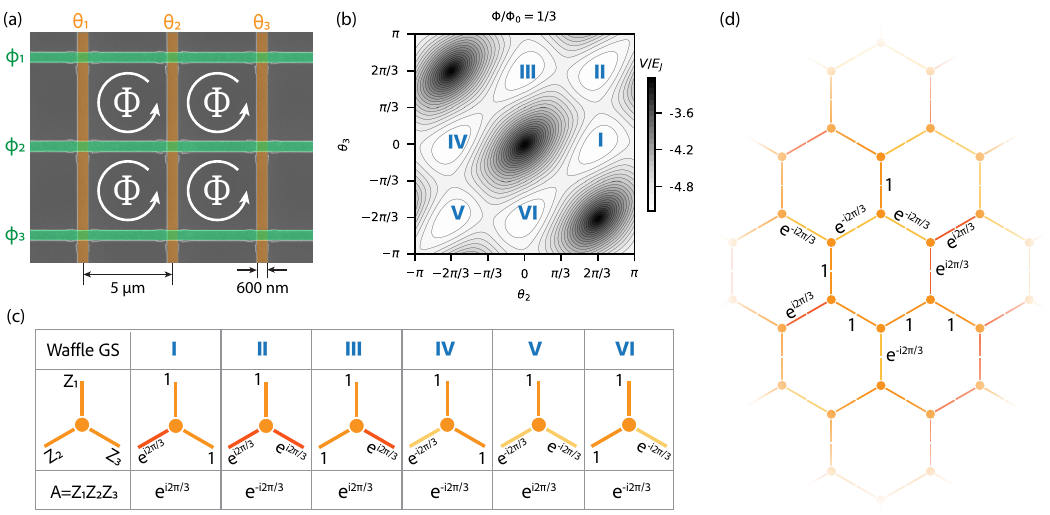}
    \caption{\textbf{Crossbar array and gauge symmetries.} (a) False-colored scanning electron micrograph of a ``waffle'' crossbar array. 
    3 gauge wires (orange, $\theta_{1,2,3}$) cross 3 matter wires (green, $\phi_{1,2,3}$), forming 9 Josephson junctions. 
    An external magnetic flux $\Phi$ threads each of the four loops. 
    (b) Josephson potential $V$ (Eq.~\ref{eq:hj}) at $\Phi/\Phi_0=1/3$, demonstrating 6 degenerate minima in the $\theta_{2,3}$ plane.
    The plot uses the phase convention $\theta_1=0$, with $\phi_a$ adjusted to minimize the energy at each point, as described in \cite{Yang2021-uj}.
    (c) In the low-energy regime, each ground state is associated with one of the 6 minima. 
    In this case $\theta_{1,2,3}$ can be treated as clock operators $Z_i$. 
    The second row shows the eigenvalues of $Z_{1,2,3}$ when acting on each of the 6 ground states. 
    The third row shows the eigenvalue of the star operator $A=Z_1\,Z_2\,Z_3$ when acting on each ground state, showing that they satisfy Eq.~\ref{eq:h_star}. 
    (d) The full $\mathbb{Z}_3$ lattice gauge theory is constructed by tiling waffles (each represented by 3 gauge spins) into a hexagonal lattice. 
    A pair of neighboring vertices shares one gauge spin. The phases on the shared spins (legs) illustrate a sample configuration of minima on the full lattice.}
    \label{fig:z3}
\end{figure*}

Here, to resolve the 6 degenerate minima in the Josephson potential, we overcome the limitations of transport measurements by employing a circuit quantum electrodynamics probe \cite{Blais2021-hi}.
We capacitively couple one gauge wire of the waffle to a $\lambda/2$ readout resonator and perform microwave spectroscopy as a function of magnetic flux.
Observations are in excellent agreement with spectra calculated using a neural-network variational Monte Carlo (NN-VMC) approach.
By further examining fine structure in the excitation spectrum, as well as external symmetry breaking imposed by the readout resonator, we gain direct evidence for the desired six‑well potential landscape.
These results demonstrate that the required $\mathbb{Z}_3$ CGS can be faithfully realized in a superconducting circuit, establishing the waffle as a building block for quantum spin liquids, and promoting crossbar junction arrays as a promising platform for quantum simulation in general.

\section{Circuit concept}

The central element of our circuit is the waffle (Fig.~\ref{fig:z3}a).
It is composed of three vertical superconducting wires with phases $\theta_i$ ($i=1,2,3$) and three horizontal superconducting wires with phases $\phi_a$ ($a=1,2,3$), coupled at every crossing by nominally identical Josephson junctions of energy $E_J$, yielding a potential energy of the form 
\begin{equation}
\label{eq:hj}
V = -E_J \sum_{i,a} \bigl[ W_{ai}(\Phi)\, e^{i(\theta_i-\phi_a)} + \text{h.c.} \bigr].
\end{equation}
A static magnetic flux $\Phi$ threads each elementary loop, and is encoded in the coupling matrix $W(\Phi)$. At $\Phi=\Phi_0/3$, which we refer to as the CGS point, the system exhibits an enhanced symmetry consisting of $\mathbb{Z}_3$ phase shifts $\theta_i \to \theta_i \pm 2\pi/3$ in the gauge wires, accompanied by permutations of the phases $\phi_a$ in the matter wires (hence the ``combinatorial'' in CGS).
At the CGS point the potential has six degenerate minima (Fig.~\ref{fig:z3}b). See Supplement Sec.~\ref{sec: symmetries} for details.

Remarkably, the degenerate ground states satisfy a $\mathbb Z_3$ star-product rule.  
Denoting $Z_i=e^{i\theta_i}$ at the potential minima -- with eigenvalues $1$, $e^{i2\pi/3}$, and $e^{-i2\pi/3}$ -- the star operator $A = Z_1 Z_2 Z_3$ (Fig.~\ref{fig:z3}c) satisfies $A=e^{\pm i 2 \pi / 3}$ in the six ground states.
The effective low-energy Hamiltonian at $\Phi=\Phi_0/3$ can therefore be written as
\begin{equation}
    \label{eq:h_star}
    H_{\rm star} = J (A + A^\dagger),
\end{equation}
where $J = (\sqrt{3}-1)E_J$.
This star-product constitutes a crucial ingredient for building a spin-liquid ground state when multiple waffles are interconnected to form a lattice (Fig.~\ref{fig:z3}d).
Another ingredient, a plaquette operator, is provided by tunneling between symmetric ground states \cite{Yang2021-uj}.

\begin{figure}
    \centering
    \includegraphics[width=1\linewidth]{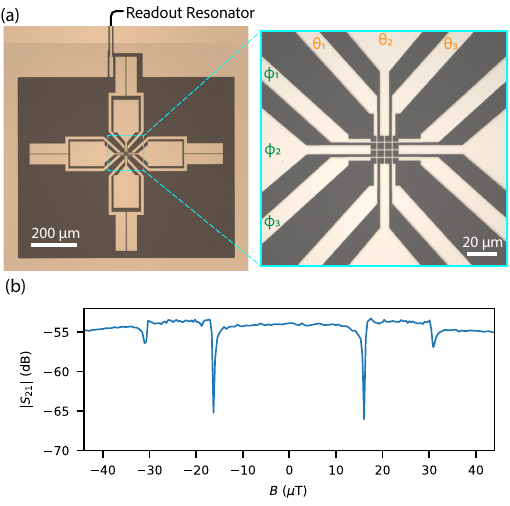}
    \caption{\textbf{Measurement architecture.}
    (a) Optical micrograph showing capacitor pads connected to gauge wires ($\theta_{1,2,3}$) and matter wires ($\phi_{1,2,3}$), designed to respect combinatorial gauge symmetry in the $6\times6$ capacitance matrix.
    Left gauge wire $\theta_1$ is weakly coupled to a $\lambda/2$ readout resonator. 
    (b) Modulation of the transmitted signal $S_{21}$ near resonance ($f_\mathrm{signal}=6.87\,\text{GHz}$) by the applied external magnetic field $B$. 
    Dips are due to hybridization between the waffle eigenmodes and the readout resonator.
    }
    \label{fig:device}
\end{figure}

The goal of the current investigation is to determine whether the symmetry structure discussed above -- six degenerate minima satisfying Eq.~\ref{eq:h_star} -- can be realized in experiment.
We therefore focus on the simplest setting where tunneling between Josephson minima is weak.
To reach this limit, we add large capacitance pads to each gauge and matter wire to suppress the charging energy -- an approach analogous to that used in transmon qubits (Fig.~\ref{fig:device}a).
A key distinction between the CGS and earlier approaches \cite{Ioffe2002-aa,Duocot2003-aa} is that CGS is non-perturbative, meaning it is not broken by quantum fluctuations.
This feature is ensured if the $6\times6$ capacitance matrix $C$ is invariant under permutations of the matter wires \cite{Chamon2020-oz,Yang2021-uj}, which we verify to within $6\%$ using finite‑element simulations (Supplement Sec.~\ref{sec:cap}).

For readout, the waffle is embedded in a circuit‑QED architecture (Fig.~\ref{fig:device}a).
A $\lambda/2$ coplanar‑waveguide resonator with a bare resonant frequency $f_{R,\text{bare}}=6.872~\mathrm{GHz}$ is capacitively coupled to the first gauge wire ($\theta_1$) via a small coupling capacitance $C_\mathrm{couple}\approx 7.5\,\text{fF}$.
The resonator is also coupled to a $50\,\Omega$ feedline with external quality factor $Q_c \approx 30\,000$ (Fig.~\ref{fig:res}).
The resonator acts as a sensitive dispersive probe: the effective admittance of the waffle modifies the resonator’s resonance frequency, a change that we detect by measuring the complex microwave transmission $S_{21}$ through the feedline.
Further details on the fabrication of the waffle and the readout resonator are provided in Supplement Sec.~\ref{sec:fab}.

\section{Results}

An external out‑of‑plane magnetic field $B$ tunes the flux $\Phi$ per loop and thereby controls the coupling matrix $W(\Phi)$ and the potential landscape.
We first probe the waffle with a single‑tone measurement.
The microwave signal is fixed near the bare resonator frequency, and the complex transmission $S_{21}$ is recorded while sweeping $B$.
Measured transmission shows a series of sharp dips (Fig.~\ref{fig:device}b), which arise from hybridization of the readout resonator with the waffle’s collective modes.
To map out the full spectrum we use two‑tone spectroscopy.
A weak probe tone is kept at the resonator frequency to continuously monitor $S_{21}$, while a second pump tone is injected through the same feedline and its frequency is swept.

\begin{figure*}[!t]
    \centering
    \includegraphics[width=1\linewidth]{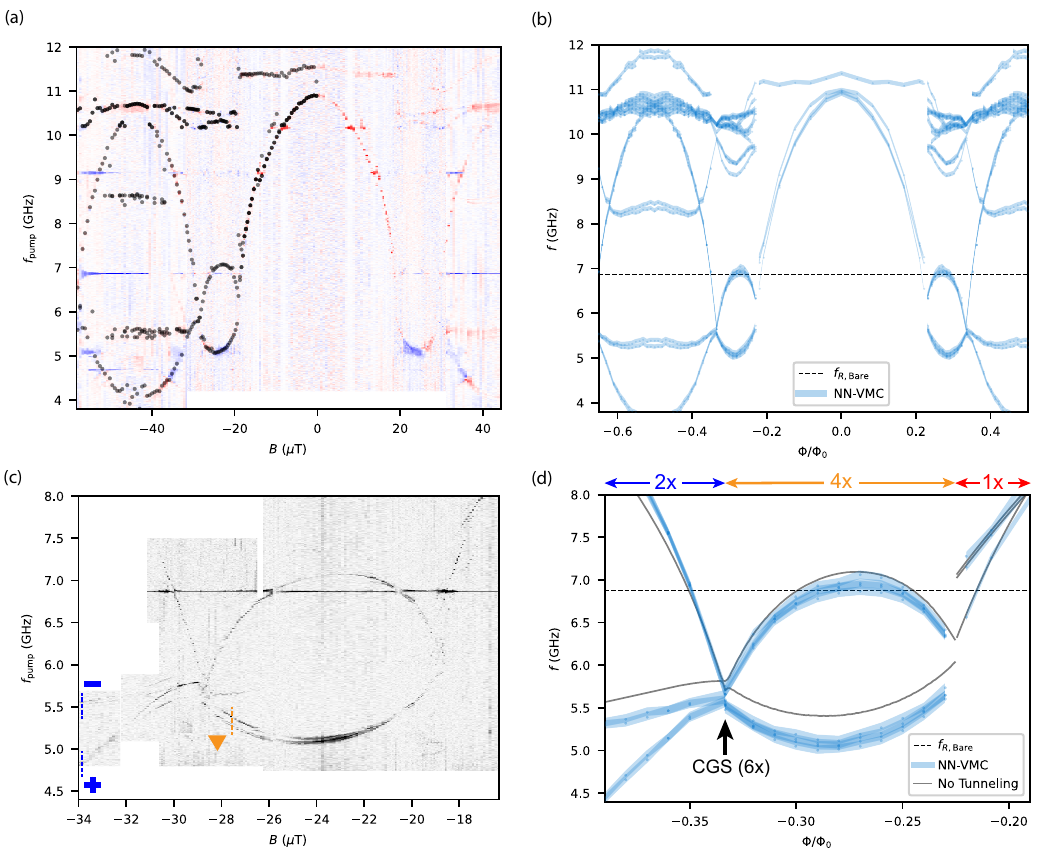}
    \caption{\textbf{Two-tone spectroscopy.} (a) Phase response of transmitted signal as a function of pump frequency $f_\mathrm{pump}$ and external magnetic field $B$.
    Contrast is adjusted at each flux value to maximize the visibility of the response. 
    Black dots indicate extracted peaks for $B<0$.
    (b) Transition frequencies of the waffle calculated by NN-VMC, using ideal CGS conditions. 
    Blue band represents uncertainty estimated using a block jackknife method (Supplement Sec.~\ref{sec:NNVMC}).
    The dashed line indicates the bare resonance of the readout resonator.
    (c) Zoomed-in two-tone spectroscopy (amplitude response) near the critical CGS field ($B=-28.7\,\mu\text{T}$). 
    (d) Zoomed-in NN-VMC spectrum near the critical CGS flux ($\Phi=-\Phi_0/3$).
    Also shown is the spectrum calculated analytically within a harmonic approximation (black solid line), which discards the effect of tunneling between minima as well as the anharmonicity of each minimum.
    Notice that the harmonic approach misses a band of transitions below the CGS point ($\Phi/\Phi_0<-1/3$), which is related to tunneling.
    Arrows indicate the number of degenerate minima at different fluxes.
    As flux is decreased the system evolves from one (red), to four (orange), to two (blue) minima.
    At the CGS point there are six degenerate minima (black).
    }
    \label{fig:wide}
\end{figure*}

Figure~\ref{fig:wide}a shows the measured two‑tone spectrum as a function of the external magnetic field $B$ and the pump frequency $f_{\text{pump}}$. 
The observed features have a periodicity of $\Delta B \approx 86\,\mu\text{T}$, close to the expected period of $80\,\mu\text{T}$ for our loop area, which confirms that the response faithfully tracks the flux $\Phi$ threading the waffle. 
At zero magnetic flux the spectrum is featureless below $10~\mathrm{GHz}$, qualitatively reflecting the presence of a single, stiff minimum with all $\theta_i = \phi_a = 0$.
As the magnetic field is increased, transitions move down in frequency and split in a complex pattern, reflecting the softening and bifurcation of the waffle potential into a complex, multi-minima landscape (Supplement Sec.~\ref{sec: landscape}).

To interpret the spectroscopy more precisely, 
we compute the expected spectrum of an ideally symmetric waffle by constructing neural networks to generate variational states sampled via quantum Monte Carlo (NN-VMC). 
Neural networks are uniquely suited to this problem: 
their expressive variational Ans\"atze can capture the complexity of high-dimensional quantum states in ways that traditional methods cannot. 
The degrees of freedom of the six wires map onto a quantum particle moving in a potential in a six-dimensional space, 
or equivalently, six particles on a ring (Supplement Sec.~\ref{sec:NNVMC}). 
Our NN-VMC method for superconducting circuits builds on recent approaches for excited-state calculations that circumvent explicit orthogonalization by solving an auxiliary ground-state problem~\cite{Pfau2024NES}.

The neural-network-determined spectra reproduce all major features observed in the experiment, including the initial lowering of spectral features as $B$ is increased from zero, subsequent splitting and non-monotonic field evolution, and flat spectral features that emerge over an isolated range of magnetic flux (Fig.~\ref{fig:wide}b).
Quantitative discrepancies between experiment and theory likely originate from differences between the experimental parameters, such as the Josephson and charging energies, and those assumed in theory.

Near the CGS point, low-lying spectral features form a characteristic dome structure (Fig.~\ref{fig:wide}c), a signature of the system's evolution between one, four, six, and two degenerate minima (Fig.~\ref{fig:wide}d).
Bands of low-lying transitions cross at the six-minima CGS point (black arrow in Fig.~\ref{fig:wide}d), which is on the boundary between the four-minima region at $\Phi/\Phi_0>-1/3$ and the two-minima region at $\Phi/\Phi_0<-1/3$.
Comparing the NN-VMC spectrum with a harmonic-approximation calculation (Supplement Sec.~\ref{sec:harmonic}) that neglects inter-well tunneling, 
we infer that the two low-lying bands below the CGS flux ($\Phi/\Phi_0<-1/3$) arise from tunneling between the two degenerate minima (Fig.~\ref{fig:wide}d): 
without tunneling the harmonic approximation predicts only a single band there. 
These two tunneling-split bands are also resolved in experiment, and are marked by blue dashed lines in Fig.~\ref{fig:wide}c.
Overall, the excellent qualitative agreement between the measured and calculated spectra provides strong evidence that the desired Josephson potential $V$ (Eq.~\ref{eq:hj}), along with a properly symmetrized capacitance, has been realized in our device.

We now turn to fine structure in the waffle spectra, which provides further evidence for the predicted configuration of minima. 
The physical picture is that weak tunneling or symmetry breaking, for instance from disorder or coupling to the readout resonator, splits the degeneracies between global minima.
As a result, there is a fine-structure splitting reflecting the number of minima (Supplement Sec.~\ref{sec:finestructure}).

\begin{figure}
    \centering
    \includegraphics[width=1\linewidth]{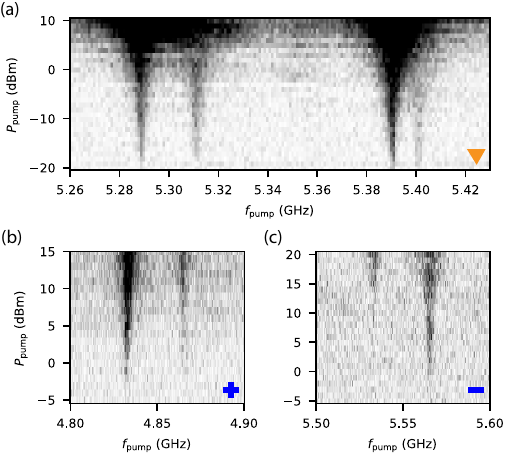}
    \caption{Two-tone spectroscopy as a function of pump frequency $f_\mathrm{pump}$ and power $P_\mathrm{pump}$ performed above (a) and below (b,c) the CGS point, as indicated by markers in Fig.~\ref{fig:wide}c.
    In the low-power limit, the mode splittings are fourfold above CGS (a) and twofold below CGS (b,c).}
    \label{fig:power}
\end{figure}

To investigate the fine-structure splitting, we performed power-dependent two-tone spectroscopy near the CGS point.
At low powers, where multi-photon transitions can be neglected, the two-tone features faithfully measure the waffle fine structure.
We focus on the lowest-lying excitations---the first excited state of each potential well---which can be split by tunneling between the wells.
At fluxes slightly above the CGS point (Fig.~\ref{fig:power}a), a quadruplet of features is observed, as expected for four minima with weak tunneling or weak symmetry breaking.
Slightly below the CGS point, the two low-lying bands are each composed of a fine-structure doublet (Fig.~\ref{fig:power}b,c), reflecting the expected twofold degeneracy from two minima.
The same fine-structure counts are also visible in the NN-VMC bands in Fig.~\ref{fig:wide}d, although the magnitude of the splittings is comparable to our numerical accuracy.
Higher-lying bands depart from expected multiplicities, and at the CGS point we are unable to perform a clear mode counting (see Supplement Sec.~\ref{sec:finestructure_exp}). 
Nonetheless, the proximity of the four‑fold and two‑fold multiplets in the low-lying bands, which converge at the critical CGS point, 
confirms the existence of six minima at the CGS point.

\begin{figure}
    \centering
    \includegraphics[width=1\linewidth]{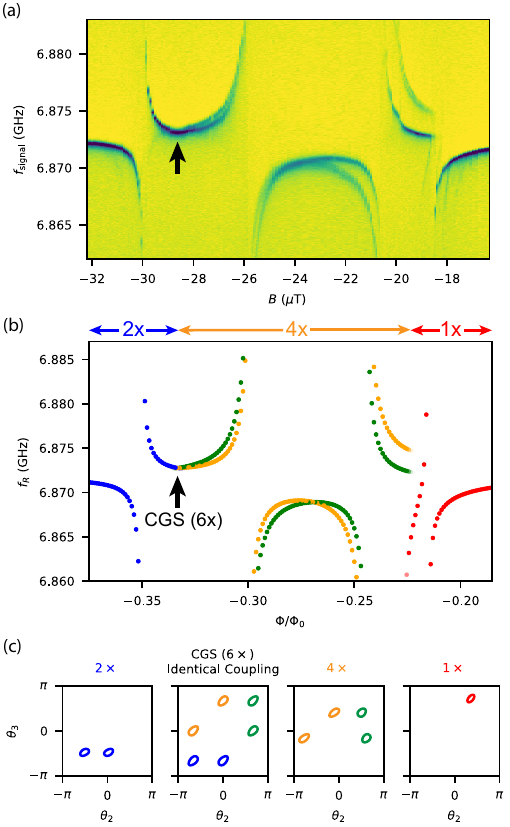}
    \caption{(a) Transmission $|S_{21}|$ of readout resonator as a function of frequency $f_\mathrm{signal}$ and magnetic field $B$. 
    Dark features are resonances, occurring at resonant frequency $f_R$.
    (b) Calculated $f_R$ as a function of magnetic flux $\Phi$, showing contributions from all global minima at a given field. 
    Colors correspond to families of minima in (c).
    Unperturbed degeneracy (1x, 4x, 6x, 2x) is indicated by arrows.
    (c) Evolution of energetic minima, shown by contours in the $\theta_2$-$\theta_3$ plane. 
    Contour colors indicate different couplings between the associated ground states and the readout resonator.
    The merging of features at the CGS point reflects the fact that the gauge ($\theta$) wires differ only by a $\mathbb Z_3$ phase.
    We emphasize that this merging is visible because spectral features are close to the bare resonator frequency at the CGS point.
    }
    \label{fig:s21}
\end{figure}

Additional evidence for CGS emerges from the hybridization of the waffle with the readout resonator, which serves as an experimentally controllable source of symmetry breaking.
The symmetry breaking is manifest as a doubling of transmission features near the waffle-resonator anticrossings (Fig.~\ref{fig:s21}a).
As shown by comparison with the harmonic-approximation spectrum (Fig.~\ref{fig:s21}b), the doubling arises from differing resonator couplings among the waffle minima, specifically in the four-minima region. 
Above the CGS point, the four populated minima fall into two subgroups of equal coupling (orange vs.\ green contours in Fig.~\ref{fig:s21}c), so the resonance appears as a doublet; 
below the CGS point, the two populated minima couple identically (blue contours in Fig.~\ref{fig:s21}c), and a single resonance remains.
Remarkably, at the critical CGS flux $\Phi = -\Phi_0/3$, the transmission evolves smoothly from a doublet into a singlet.
This convergence results from a special symmetry at the CGS point: at $\Phi=-\Phi_0/3$ the gauge-wire ($\theta$) weights in the eigenmodes are identical across all six minima (Fig.~\ref{fig:eigenvec}),
so they must couple to the readout resonator equally.
This symmetry should not come as a surprise -- it is a consequence of gauge phases being the $\mathbb{Z}_3$ clock variables at the CGS point.
Thus, the restoration of degeneracy at $\Phi=-\Phi_0/3$ is a key signature of the star Hamiltonian Eq.~\ref{eq:h_star}.

\section{Discussion and Outlook}
In summary, we have fabricated and measured a superconducting $3\times 3$ waffle circuit that realizes the local $\mathbb{Z}_3$ combinatorial gauge symmetry at the single‑vertex level. 
The observed flux‑dependent eigenmode spectrum provides direct evidence for the predicted landscape of $V$. 
Further examination of the fine structure of the eigenmodes, as well as of the waffle's coupling to the readout resonator, provides additional evidence for the existence of 6 degenerate minima at $\Phi=\pm \Phi_0/3$.
We have measured a second device, observed similar phenomena, and also measured Rabi oscillations (Fig.~\ref{fig:dv2}).

These results confirm that the key ingredient for a $\mathbb{Z}_3$ quantum double---the exact combinatorial gauge symmetry---can be engineered in a physical circuit.
The present experiment focuses on a single vertex operating in the semiclassical regime where tunneling between minima is weak.
The natural next step is to build devices deep in the quantum regime and tile many such waffles into a honeycomb lattice, where the interplay of inter‑vertex couplings and charge fluctuations should give rise to the fully interacting $\mathbb{Z}_3$ quantum double with its topologically ordered ground state. 
Beyond this specific goal, the ``waffle'' design and the underlying combinatorial gauge symmetry framework provide a flexible platform for simulating a broad class of correlated many‑body models: 
by choosing different coupling matrices, one can realize other discrete gauge groups or even graph‑based Hamiltonians with arbitrary connectivity. 
We expect that Josephson junction arrays with engineered coupling graphs will become a valuable tool in the quantum simulation of lattice gauge theories, frustrated magnets, topological phases of matter, and topological quantum computing.


\section*{Acknowledgements}
Work was supported by DOE DE-SC0026189.
H. Lo acknowledges support from the James Franck Institute Summer Research Fund.


\bibliography{main}

\onecolumngrid

\pagebreak
\setcounter{secnumdepth}{3}  
\renewcommand{\thesection}{S\Roman{section}}
\renewcommand{\thefigure}{S\arabic{figure}}
\renewcommand{\thetable}{S\arabic{table}}
\renewcommand{\theequation}{S\arabic{equation}}
\renewcommand{\thepage}{S\arabic{page}}
\setcounter{figure}{0}
\setcounter{table}{0}
\setcounter{equation}{0}
\setcounter{page}{1}
\setcounter{section}{0}

\clearpage
\begin{center}
\Large\textbf{Supporting Information for:}\\[0.5em]
\Large\textbf{\thetitle}

\vspace{1.5em}

\end{center}

\vspace{2em}

\section{Nanofabrication Recipe}
\label{sec:fab}
We first pattern and deposit the waffle on a Si substrate, following the Manhattan process: 
\begin{itemize}
  \item Clean sample: sonicate for 3 min in acetone, then 2 min in IPA, N$_2$ dry. O$_2$ plasma ash at 200 W 180 mTorr for 20 s. Dip 15 s in 10:1 BOE, DI rinse.
  \item MMA(8.5)MAA-EL13 spin 66 s at 2.2 krpm 1.5 krpm/s (film $\approx$ 600 nm), bake for 2 min at $180 \,^{\circ}\text{C}$
  \item 950PMMA-A4 spin 60 s at 2.2 krpm 1.5 krpm/s (film $\approx$ 220 nm), bake for 2 min at $180 \,^{\circ}\text{C}$
  \item Expose the bilayer resist in a Raith EBPG 5200 at 100 keV with an 880 $\mu$C/cm$^2$ dose.
  \item Cold develop in 3:1 IPA:DI at $6 \,^{\circ}\text{C}$ for 80 s with agitation
  \item Stop development with 15 s dip in IPA at $6 \,^{\circ}\text{C}$, N$_2$ dry
  \item O$_2$ plasma ash at 60 W 180 mTorr for 15 s
  \item Evaporate in Plassys MEB550S:
    \begin{itemize}
        \item Pump for 12 h to $\approx5\times10^{-8}\,\text{mbar}$; getter with Ti (3 min, 0.2 nm/s)
        \item Pressure now $\approx3\times10^{-8}\,\text{mbar}$, evaporate Al (40 nm, 1 nm/s, $\theta=-45^\circ$, $\phi=0^\circ$)
        \item Static oxidation for 40 min at 50 mbar
        \item Getter with Ti (3 min, 0.2 nm/s), pressure now $\approx3\times10^{-8}\,\text{mbar}$
        \item Evaporate Al (40 nm, 1 nm/s, $\theta=-45$, $\phi=+91^\circ$)
        \item Evaporate Al (50 nm, 1 nm/s, $\theta=-45$, $\phi=-91^\circ$)
    \end{itemize}
  \item Lift-off for 3 h in NMP at $80 \,^{\circ}\text{C}$, then sonicate for 10 min in NMP, 3 min in acetone, 2 min in IPA, N$_2$ dry.
  \item O$_2$ plasma ash at 60 W 180 mTorr for 15 s
\end{itemize}
The readout resonator, ground plane, and transmission line are then fabricated using photolithography:
\begin{itemize}
  \item AR300-80 adhesion promoter spin 45 s at 4.5 krpm, bake at $115 \,^{\circ}\text{C}$ for 2.5 min
  \item AZ MIR 703 spin 45 s at 3.5 krpm, bake at $95 \,^{\circ}\text{C}$ for 1 min
  \item Expose in a Heidelberg MLA150 with a 100 mJ/cm$^2$ dose. Post-exposure bake at $115 \,^{\circ}\text{C}$ for 1 min.
  \item Develop in AZ300-MIF for 55 s with agitation, then stop in DI for 30 s
  \item O$_2$ plasma ash at 200 W 180 mTorr for 20 s
  \item Dip 15 s in 10:1 BOE, DI rinse
  \item Evaporate Al in Plassys MEB550S at pressure $\approx3\times10^{-8}\,\text{mbar}$ (135 nm, 0.2 nm/s)
  \item Lift-off for 3 h in NMP at $80 \,^{\circ}\text{C}$, then sonicate for 10 min in NMP, 3 min in acetone, 2 min in IPA, N$_2$ dry.
  \item O$_2$ plasma ash at 60 W 180 mTorr for 15 s
\end{itemize}

\section{Capacitance Matrix of Waffle}
\label{sec:cap}
The Josephson potential energy $V$ is given by Eq.~\ref{eq:hj} in the main text, where the flux-dependent coupling matrix $W(\Phi)$ is
\begin{eqnarray}
    W(\Phi) =\frac{1}{2} \begin{pmatrix}
    1 & 1 & 1\\
    1 & e^{2\pi i \Phi/\Phi_0} & e^{4\pi i \Phi/\Phi_0}\\
    1 & e^{4\pi i\Phi/\Phi_0} & e^{8\pi i \Phi/\Phi_0}
    \end{pmatrix}.
\end{eqnarray} 
The kinetic energy $T$ of the waffle can be expressed as $T = \tfrac12 \left( \frac{\hbar}{2e} \right)^2 \bm{\dot{\varphi}}^{\mathsf T} C \bm{\dot{\varphi}}$ where $
    {\bm{\varphi}}
    =
    (\theta_1,\theta_2,\theta_3,\phi_1,\phi_2,\phi_3)^{{\mathsf T}}$
and $C$ represents the $6\times6$ capacitance matrix between the wires. A quasi-static electromagnetic simulation tool (Ansys Q3D) is used to simulate $C$. For the measured device shown in main text Fig.~\ref{fig:device}, the simulated capacitance is
\begin{eqnarray}
C_{\text{ANSYS}}\,(\text{fF})=\begin{pmatrix}
  214.77 & -44.90 & -47.49 & \textcolor{red}{-25.38} & \textcolor{red}{-25.38} & \textcolor{red}{-25.58} \\
  -44.90 & 191.28 & -44.98 & \textcolor{red}{-25.39} & \textcolor{red}{-27.00} & \textcolor{red}{-25.43} \\
  -47.49 & -44.98 & 212.40 & \textcolor{red}{-25.42} & \textcolor{red}{-25.39} & \textcolor{red}{-25.60} \\
  \textcolor{red}{-25.38} & \textcolor{red}{-25.39} & \textcolor{red}{-25.42} & \textcolor{green}{212.18} & \textcolor{orange}{-50.49} & \textcolor{orange}{-48.40} \\
  \textcolor{red}{-25.38} & \textcolor{red}{-27.00} & \textcolor{red}{-25.39} & \textcolor{orange}{-50.49} & \textcolor{green}{203.10} & \textcolor{orange}{-50.50} \\
  \textcolor{red}{-25.58} & \textcolor{red}{-25.43} & \textcolor{red}{-25.60} & \textcolor{orange}{-48.40} & \textcolor{orange}{-50.50} & \textcolor{green}{211.97}
\end{pmatrix}.
\label{eq:ansys}
\end{eqnarray}
Combinatorial gauge symmetry requires that the $C$ matrix stay invariant under permutations of the matter wires, i.e., the elements marked with the same color in Eq.~\ref{eq:ansys} (except black) should ideally have the same value. After optimizing the shape of the capacitance pads, the discrepancies in Eq.~\ref{eq:ansys} drop below $6\%$ for the current waffle design.

In order to test whether our experimental data are compatible with CGS, we perform theoretical calculations using a symmetrized capacitance matrix that respects CGS exactly,
\begin{eqnarray}
C\,(\text{fF})=\begin{pmatrix}
  214.77 & -44.90 & -47.49 & -25.5 & -25.5 & -25.5 \\
  -44.90 & 191.28 & -44.98 & -25.5 & -25.5 & -25.5 \\
  -47.49 & -44.98 & 212.40 & -25.5 & -25.5 & -25.5 \\
  -25.5 & -25.5 & -25.5 & 208 & -50 & -50 \\
  -25.5 & -25.5 & -25.5 & -50 & 208 & -50 \\
  -25.5 & -25.5 & -25.5 & -50 & -50 & 208
\end{pmatrix}.
\label{eq:cap}
\end{eqnarray}

\section{Waffle Theory}

\label{sec:theory}
We compute the waffle spectrum using two complementary theoretical descriptions. The harmonic approximation gives a semiclassical normal-mode picture by expanding the Josephson potential around each flux-dependent minimum, while NN-VMC provides a variational quantum calculation of the low-lying eigenstates on the five-dimensional compact torus obtained after removing the global phase mode. The comparison between these approaches identifies which spectral features can be understood as local oscillations within individual wells, and which require tunneling and anharmonicity beyond the harmonic approximation.

\subsection{Waffle Hamiltonian and its Symmetries}
\label{sec: symmetries}
We start from the classical Lagrangian {in the presence of magnetic flux $\Phi$ per plaquette, with kinetic and potential terms given by}
\begin{equation}
    {\mathcal L}
    =
    \frac{1}{2}\left(\frac{\hbar}{2e}\right)^2{\bm{\dot{\varphi}}^{\mathsf T}}C{\bm{\dot{\varphi}}}
    -
    V(\bm{\varphi};\Phi),
    \label{eq:lagrangian_6d}
\end{equation}
where, {as before,} $
    {\bm{\varphi}}
    =
    (\theta_1,\theta_2,\theta_3,\phi_1,\phi_2,\phi_3)^{{\mathsf T}}$
collects the superconducting phases of the matter and gauge wires, and $C$ is taken from Eq.~\ref{eq:cap}. Based on the Lagrangian in Eq.~\ref{eq:lagrangian_6d}, the canonical charge operator is given by
\begin{equation}
    {\bm{q}}=\nabla_{{\dot{\bm \varphi}}} {\mathcal{L}}=\left(\frac{\hbar}{2e}\right)^2C{\dot{\bm \varphi}}
    \;.
\end{equation}
The Hamiltonian is then given by
\begin{equation}
    H={\bm{q}^{\mathsf{T}} \dot{\bm \varphi}}-{\mathcal L}=\frac{2e^2}{\hbar^2}{\bm q^{\mathsf{T} }}C^{-1}{\bm{q}}+V({\bm{\varphi}};\Phi)
    \;.\label{eq: quantum Hamiltonian}
\end{equation}
Upon canonical quantization, the {components} ${q}_\mu=-i\hbar\partial/\partial{{\phi}_{\mu}}$ {for $\mu=0, \ldots, 5$} satisfy the charge-phase commutation relations
\begin{equation}
    {[\varphi_\mu, q_\nu]}=i\hbar\delta_{\mu\nu}
    \;.
    \label{eq:chage-phase commutation}
\end{equation}

The Josephson potential obtained from $ V$ in Eq.~\ref{eq:hj}
\begin{eqnarray}
    V(\bm \varphi, \Phi) = -E_J \, \bm {e^{i \varphi}}\, \bm {W}(\Phi)\,  \bm{e^{-i \varphi}}\quad \text{with}\quad \bm W(\Phi) = 
    \begin{pmatrix}
        0 &  W(\Phi)\\
        W^\star(\Phi) & 0
    \end{pmatrix},
    \label{eq: V}
\end{eqnarray}
is invariant under the global $U(1)$ shift $ \varphi_\mu \rightarrow  \varphi_\mu +\alpha$. 
At the CGS point, the coupling matrix $  \bm W_{\mathrm{CGS}} \equiv \bm W(\Phi_0/3)$ possesses additional discrete symmetries. These symmetries act on the phase variables as $\bm{e^{i \varphi}}  \rightarrow \bm M\, \bm{e^{i \varphi}}$ (accompanied by an appropriate transformation of the charge operators)
so that they preserve both the Hamiltonian Eq.~\ref{eq: quantum Hamiltonian} and the algebra in Eq.~\ref{eq:chage-phase commutation}. 
The symmetry $\bm M$ is generated by  $\mathbb Z_2$ complex conjugations $\bm S$ and $\mathbb Z_3$ phase shifts/permutations $\bm R$, explicitly given by
\begin{eqnarray}
    \bm S =\begin{pmatrix}
        1 & 0\\
        0 & P_{(23)}
    \end{pmatrix}K\quad \text{and}\quad  \bm{R} = \begin{pmatrix}
        Z& 0\\
        0 & P_{(132)}
    \end{pmatrix}.
\end{eqnarray} 
 Here $K$ denotes complex conjugation, $
 Z = \mathrm{diag}(1,e^{2\pi i/3}, e^{-2\pi i/3})$, and $P_\pi$ is the $3\times 3$ permutation matrix associated with $\pi \in S_3$. These generators do not commute. Instead, they satisfy
 \begin{eqnarray}
     \bm S^2=1, \quad \bm R^3=1, \quad \text{and}\quad \bm S\, \bm R\, \bm S = \bm R^{-1}.
 \end{eqnarray}
These relations show that $\mathbb Z_2$ acts non-trivially on $\mathbb Z_3$, so the symmetry group is $S_3 \cong \mathbb Z_3\rtimes \mathbb Z_2$. Accordingly, the six degenerate ground states shown in Fig.~\ref{fig:z3}b are naturally identified with the elements of the symmetry group
$S_3$ and may be labeled as
\begin{align}
\mathrm{I}: 1,\qquad
\mathrm{II}: \bm S\, \bm R,\qquad
\mathrm{III}: \bm R^2,\qquad
\mathrm{IV}: \bm S,\qquad
\mathrm{V}: \bm R,\qquad
\mathrm{VI}: \bm S\,\bm R^2.
\end{align}

\subsection{Separation of the Center-of-Mass Mode}
The Josephson potential is a function only of the phase differences between the two sets of wires, i.e. of differences of the form $\theta_i-\phi_a$. Therefore, it is independent of the center-of-mass mode
\begin{equation}
    \Xi
    =
    \frac{1}{\sqrt{6}}
    \left(
    \sum_{i=1}^{3}\theta_i
    +
    \sum_{a=1}^{3}\phi_a
    \right).
\end{equation}
We introduce an orthogonal transformation $U$
\begin{equation}
    (\Xi,\bm \xi)^{{\mathsf T}}=U^{{\mathsf T}}{\bm \varphi} \quad {\text{and}\quad (N, \bm n)^{\mathsf T} = U^{\mathsf T}  \bm q},
    \label{eq: canonical}
\end{equation}
with
\begin{equation}
     U=[u_0,u_1,\cdots,u_5]=
     \begin{pmatrix}
     1/\sqrt{6} & 1/\sqrt{2} & 1/\sqrt{6} & 0 & 0 & 1/\sqrt{6} \\
     1/\sqrt{6} & -1/\sqrt{2} & 1/\sqrt{6} & 0 & 0 & 1/\sqrt{6} \\
     1/\sqrt{6} & 0 & -2/\sqrt{6} & 0 & 0 & 1/\sqrt{6} \\
     1/\sqrt{6} & 0 & 0 & 1/\sqrt{2} & 1/\sqrt{6} & -1/\sqrt{6} \\
     1/\sqrt{6} & 0 & 0 & -1/\sqrt{2} & 1/\sqrt{6} & -1/\sqrt{6} \\
     1/\sqrt{6} & 0 & 0 & 0 & -2/\sqrt{6} & -1/\sqrt{6}
     \end{pmatrix}.
 \label{eq:U_transform}
\end{equation}
In this basis the capacitance matrix becomes
\begin{equation}
    C'
    =
    U^{{\mathsf T}}CU
    =
    \begin{pmatrix}
        c & d^{{\mathsf T}}\\
        d & D
    \end{pmatrix}.
    \label{eq:Cprime_block}
\end{equation}

Although the center-of-mass mode has been removed from the potential, it can still couple to the relative modes through the capacitance matrix. To eliminate this kinetic coupling, we perform a Gram-Schmidt transformation with respect to the capacitance metric. Let $u_0$ denote the center-of-mass mode column of $U$, and let $u_i$ with $i=1,\ldots,5$ denote the relative mode columns. We define
\begin{equation}
    \tilde u_i
    =
    u_i
    -
    u_0\frac{C'_{0i}}{C'_{00}},
    \qquad i=1,\ldots,5,
    \label{eq:gram_schmidt}
\end{equation}
and construct {the non-orthogonal transformation}
\begin{equation}
    \tilde U
    =
    [u_0,\tilde u_1,\ldots,\tilde u_5]
\end{equation}
and its inverse
\begin{equation}
    \tilde U^{-1}
    =
    [\tilde u_0, u_1,\ldots, u_5]^{{\mathsf T}}
    \;,
\end{equation}
where 
\begin{equation}
\tilde u_0
    =
    u_0
    + \sum_{j=1}^5
    u_j\frac{C'_{0j}}{C'_{00}}
    \;.
\end{equation}
Under such a transformation, the coordinates (phases) become
\begin{equation}
    (\tilde \Xi,\tilde{\bm \xi})^{{\mathsf T}}
    =
    \tilde U^{-1}{\bm \varphi} ,
    \label{eq:phase tf}
\end{equation}
while the charge operators $\tilde N=-i\partial/\partial_{\tilde\Xi}$ and $\tilde{\bm n}=-i\bm\nabla_{\bm\tilde{\xi}}$ transform as
\begin{equation}
    (\tilde N,\tilde{\bm{n}})^{{\mathsf T}}
    =
    \tilde U^{{\mathsf T}}{\bm q} ,
    \label{eq:charge tf}
\end{equation}
Then by Eq. \ref {eq:phase tf} and Eq. \ref{eq:charge tf}, the phase-charge commutation relations Eq.~(\ref{eq:chage-phase commutation}) are still preserved
\begin{equation}
    [\tilde{\Xi},\tilde{N}]=i\hbar,\quad [\tilde{\xi}_i,\tilde{n}_j]=i\hbar\delta_{ij}.
\end{equation}
{In fact, $\tilde U$ corresponds to a linear canonical transformation, preserving the commutation relations.}

Comparing Eqs.~\ref{eq: canonical} and \ref{eq:phase tf}, the new coordinates are related to the original orthogonal coordinates by
\begin{equation}
    \tilde \Xi
    =
    \Xi
    +
    \sum_{i=1}^{5}
    \frac{C'_{0i}}{C'_{00}}\xi_i,
    \qquad
    \tilde \xi_i=\xi_i.
    \label{eq:q_tilde_relation}
\end{equation}
The Josephson potential remains independent of $\tilde \Xi$. Indeed, for any phase difference entering the Josephson energy,
\begin{equation}
    \varphi_\mu-\varphi_\nu
    =
    (\tilde U_{\mu0}-\tilde U_{\nu0})\tilde \Xi
    +
    \sum_{k=1}^{5}
    (\tilde U_{\mu k}-\tilde U_{\nu k})\tilde \xi_k, \quad {\text{for}\quad \mu, \nu=0, \ldots, 5. }
\end{equation}
Since the center-of-mass mode column has identical entries, $\tilde U_{\mu0}=\tilde U_{\nu0}=1/\sqrt{6}$, the first term vanishes:
\begin{equation}
    \varphi_\mu-\varphi_\nu
    =
    \sum_{k=1}^{5}
    (\tilde U_{\mu k}-\tilde U_{\nu k})\tilde \xi_k .
\end{equation}
Thus the potential depends only on the five relative coordinates. Since $\tilde{\bm \xi}=\bm \xi$, we use $\bm \xi$ in the following.

The capacitance matrix now becomes block-diagonal in the new basis
\begin{equation}
    \tilde C
    =
    \tilde U^{{\mathsf T}}C\tilde U
    =
    \begin{pmatrix}
        c & 0\\
        0 & C_{\rm rel}
    \end{pmatrix},
    \qquad
    C_{\rm rel}
    =
    D-\frac{dd^{{\mathsf T}}}{c}.
    \label{eq:Ctilde_block}
\end{equation}
The inverse transforms as
\begin{equation}
    \tilde C^{-1}
    =
    \tilde U^{-1}C^{-1}\tilde U^{{-\mathsf T}}
    =
    \begin{pmatrix}
        g_{\tilde{\Xi}} & 0\\
        0 & G_\xi
    \end{pmatrix},
    \qquad
    g_{\tilde{\Xi}}=1/c, 
    \qquad
    G_\xi=C_{\rm rel}^{-1}
    \label{eq:Ctilde_inv_block}
\end{equation}
Thus, the Hamiltonian separates into the center-of-mass mode and relative-mode parts:
\begin{equation}
    H
    =
    -\frac{2e^2}{\hbar^2}g_{\tilde{\Xi}}\partial_{\tilde \Xi}^{2}
    -
    \frac{2e^2}{\hbar^2}\bm\nabla_{\bm\xi}^{{\mathsf T}}G_\xi\bm\nabla_{\bm\xi}
    +
    V(\bm{\xi};\Phi),
    \label{eq:H_Q_q}
\end{equation}
and the wavefunction can be written as
\begin{equation}
    \psi(\tilde \Xi,\bm\xi)
    =
    e^{i\tilde{N}\tilde \Xi/\hbar}\chi(\bm\xi).
\end{equation}
For the low-energy sector we take $\tilde{N}=0$, reducing Eq.~\ref{eq:H_Q_q} to the five-dimensional Hamiltonian
\begin{equation}
    H_\xi
    =
    -\frac{2e^2}{\hbar^2}\bm\nabla_{\bm \xi}^{{\mathsf T}}G_\xi\bm\nabla_{\bm\xi}
    +
    V(\bm\xi;\Phi).
    \label{eq:Hq}
\end{equation}

\subsection{Calculating Waffle Eigenmodes Using Harmonic Approximation}
\label{sec:harmonic}
At a certain flux $\Phi$, we find all minima of the Josephson potential energy $V(\bm\varphi;\Phi)$ defined in Eq.~\ref{eq: V}.
Near each minimum, we expand $V(\bm\varphi;\Phi)$ to $2^{\text{nd}}$ order in the harmonic approximation: $V\sim V_0+\frac{1}{2} \left( \frac{\hbar}{2e} \right)^2 \bm{\varphi}^{\mathsf T} L^{-1} \bm{\varphi}$, where the elements of the inverse inductance matrix are:
\begin{eqnarray}
\left( \frac{\hbar}{2e} \right)^2 (L^{-1})_{\mu\nu}=\frac{\partial V}{\partial\varphi_\mu\partial\varphi_\nu}.
\end{eqnarray}
Each minimum is then treated as an individual harmonic oscillator, while tunneling between different minima is not accounted for. For each minimum we calculate its normal modes, i.e., the transition frequencies between its ground and excited states. Under the harmonic approximation, we can write the Lagrangian in Eq.~\ref{eq:lagrangian_6d} in quadratic form:
\begin{equation}
\mathcal{L}_\mathrm{HA} =\frac{1}{2} \left( \frac{\hbar}{2e} \right)^2 \bm{\dot{\varphi}}^{\mathsf T} C \bm{\dot{\varphi}} - \frac{1}{2} \left( \frac{\hbar}{2e} \right)^2 \bm{\varphi}^{\mathsf T} L^{-1} \bm{\varphi},
\label{eq: Lquadratic}
\end{equation}
The eigenmodes form a basis that simultaneously diagonalizes both $C$ and $L^{-1}$ into $\widetilde{C}$ and $\widetilde{L^{-1}}$, respectively:
\begin{eqnarray*}
    \widetilde{C} &=& P^{\mathsf T} C P \\
    \widetilde{L^{-1}} &=& P^{\mathsf T} L^{-1} P\,.
\end{eqnarray*}
The matrix $P=[p_0,p_1,\cdots, p_5]$ has eigenvectors $p_\mu$ as columns, and it transforms $\bm{\varphi}$ into the eigenbasis representation by $\widetilde{\bm{\varphi}}=P^{-1}\bm{\varphi}$. Thus, the Lagrangian can be expressed as:
\begin{equation}
\mathcal{L}_\mathrm{HA} = \sum_{\mu=0}^{5} \left[\frac{1}{2} \left( \frac{\hbar}{2e} \right)^2 \widetilde{C}_{\mu\mu}\, \dot{\widetilde{\varphi}}_\mu^2 - \frac{1}{2} \left( \frac{\hbar}{2e} \right)^2 (\widetilde{L^{-1}} )_{\mu\mu}\, {\widetilde{\varphi}}_\mu^2\right].
\end{equation}
From $\mathcal{L}$ we can extract the canonical charge in the eigenbasis, $\widetilde{q}_\mu=\partial\mathcal{L}/\partial\dot{\widetilde{\varphi}_\mu}=\left( \frac{\hbar}{2e} \right)^2 \widetilde{C}_{\mu\mu}\,\dot{\widetilde{\varphi}}_\mu$, yielding the desired Hamiltonian:
\begin{equation}
    H_{\rm HA}=\sum_{\mu=0}^5\widetilde{q}_\mu\,\dot{\widetilde{\varphi}}_\mu - \mathcal{L}_\mathrm{HA}=\sum_{\mu=0}^{5} \left[\frac{1}{2} \left( \frac{2e}{\hbar} \right)^2 \frac{\widetilde{q}_\mu^2}{\widetilde{C}_{\mu\mu}} + \frac{1}{2} \left( \frac{\hbar}{2e} \right)^2 (\widetilde{L^{-1}} )_{\mu\mu}\, {\widetilde{\varphi}}_\mu^2\right]
\end{equation}
The above Hamiltonian has 6 eigenmodes with energies $\hbar\,\omega_\mu=hf_\mu=\sqrt{(\widetilde{L^{-1}} )_{\mu\mu}/\widetilde{C}_{\mu\mu}}$. Due to the global $U(1)$ symmetry, one of the frequencies vanishes, $\omega_0=0$, leaving us with 5 nontrivial eigenmodes $\omega_i$ with $i=1,\ldots, 5$. The phase $\widetilde{\varphi}_i$ and its conjugate momentum $\widetilde{q}_i$ obey the canonical commutation relations $[\tilde{\varphi_i}, \tilde{q_j}]=i\hbar\delta_{ij}$. The lowering ladder operator for mode $\omega_i$ is given by:
\begin{equation}
    \hat{a}_i = i \frac{1}{\sqrt{2 \widetilde{C}_{ii}\,\hbar\,\Omega_i}}\, \tilde{q_i} + \sqrt\frac{(\widetilde{L^{-1}} )_{ii}}{2\, \hbar\,\Omega_i}\,\tilde{\varphi_i}\,,
\end{equation}
which enables us to write the full Hamiltonian in terms of the raising and lowering operators as
\begin{equation}
H_{\rm HA}=\sum_{i=1}^5\hbar\,\omega_i\left(\hat{a}_i^\dagger\hat{a}_i+\frac{1}{2}\right)\,.
\end{equation}

\subsection{Variational Monte Carlo Calculations}
\label{sec:NNVMC}
To obtain the transition frequencies measured in two-tone spectroscopy, we compute several low-lying eigenstates of the effective five-dimensional waffle Hamiltonian Eq.~\ref{eq:Hq}. Each state is represented by a complex neural-network wavefunction defined on the compact torus, with periodic input features chosen to respect the phase periodicity of the superconducting circuit.

{Because the spectrum requires multiple excited states rather than only the ground state, we build the natural excited states (NES)~\cite{Pfau2024NES} as our variational ansatz.} In this approach, several neural-network wavefunctions are optimized simultaneously as a single enlarged variational object. The construction prevents the different variational states from collapsing onto the same state, without introducing explicit orthogonalization or overlap-penalty terms. After optimization, the state-resolved energies are obtained by diagonalizing the averaged energy matrix (see Ref.~\cite{Pfau2024NES} for details) within the optimized low-energy subspace. The resulting low-energy transition frequencies are compared with the experimentally measured two-tone spectrum in main text Fig. \ref{fig:wide}. 

In the original {$\bm \varphi$} basis, all the six phases are naturally periodic over $2\pi$. After the phase transformation, we need to determine the new periodic patterns of the five-dimensional model. The Josephson potential in the $\bm\xi$ coordinates can be written as 
\begin{equation}
    V(\bm\xi;\Phi)
    =
    -
   E_J
    \sum_{m=1}^{9}
    \cos\!\left[(K_\xi \bm\xi)_m+\delta_m\right],
    \label{eq:Vq}
\end{equation}
where {$K_\xi$ is a $9\times 5$ coupling matrix  and} the flux offsets are collected in
\begin{equation}
    \bm\delta/2\pi
    =
    (0,0,0,0,-\Phi/\Phi_0,-2\Phi/\Phi_0,0,-2\Phi/\Phi_0,-4\Phi/\Phi_0)^{{\mathsf T}} .
\end{equation}
We choose the normalization of $K_\xi$ such that
\begin{equation}
    K_\xi\Sigma=K_{\rm int},
    \qquad
    \Sigma
    =
    {\rm diag}(\sqrt{2},\sqrt{6},\sqrt{2},\sqrt{6},\sqrt{6}),
\end{equation}
with
\begin{equation}
    K_{\rm int}
    =
    \begin{pmatrix}
        1 &  1 & -1 & -1 & 2\\
       -1 &  1 & -1 & -1 & 2\\
        0 & -2 & -1 & -1 & 2\\
        1 &  1 &  1 & -1 & 2\\
       -1 &  1 &  1 & -1 & 2\\
        0 & -2 &  1 & -1 & 2\\
        1 &  1 &  0 &  2 & 2\\
       -1 &  1 &  0 &  2 & 2\\
        0 & -2 &  0 &  2 & 2
    \end{pmatrix}.
    \label{eq:Kint}
\end{equation}

We now introduce the $2\pi$-periodic coordinates $\bm s$ through
\begin{equation}
    \bm\xi
    =
    \Sigma A^{-1}\bm s
    \equiv
    W\bm s,
    \label{eq:q_to_s}
\end{equation}
where
\begin{equation}
     A=
	\begin{pmatrix}
		2 & 0 & 0 & 0 & 0\\
		-1 & 3 & 0 & 0 & 0\\
		0 & -2 & -1 & -1 & 2\\
		0 & 0 & 2 & 0 & 0\\
		0 & 0 & -1 & 3 & 0
	\end{pmatrix}.
    \label{eq:A_basis}
\end{equation}
Using $K_\xi\Sigma=K_{\rm int}$, the Josephson phases become
\begin{equation}
    K_\xi \xi
    =
    K_\xi\Sigma A^{-1}s
    =
    K_{\rm int}A^{-1}s
    \equiv
    N_{\rm int}s,
\end{equation}
with
\begin{equation}
    N_{\rm int}
    =
    K_{\rm int}A^{-1}
    =
    \begin{pmatrix}
        1 & 1 & 1 & 0 & 0\\
        0 & 1 & 1 & 0 & 0\\
        0 & 0 & 1 & 0 & 0\\
        1 & 1 & 1 & 1 & 0\\
        0 & 1 & 1 & 1 & 0\\
        0 & 0 & 1 & 1 & 0\\
        1 & 1 & 1 & 1 & 1\\
        0 & 1 & 1 & 1 & 1\\
        0 & 0 & 1 & 1 & 1
    \end{pmatrix}
    \in \mathbb{Z}^{9\times5}.
    \label{eq:Nint}
\end{equation}
Therefore the potential becomes
\begin{equation}
    V(\bm s)
    =
    -
    E_J
    \sum_{m=1}^{9}
    \cos\!\left[(N_{\rm int}\bm s)_m+\delta_m\right].
    \label{eq:Vs}
\end{equation}
Since $N_{\rm int}$ has integer entries, $V(s)$ is manifestly $2\pi$-periodic in each component of $s$.

Finally, because $\bm \xi=W\bm s$, the derivatives transform as
\begin{equation}
    \bm\nabla_\xi
    =
    W^{{-\mathsf T}}\bm\nabla_s .
\end{equation}
The kinetic matrix in the $s$ coordinates is therefore
\begin{equation}
    G_s
    =
    W^{-1}G_\xi W^{{-\mathsf T}}.
    \label{eq:Gs}
\end{equation}
The effective five-dimensional Hamiltonian on the torus is
\begin{equation}
    H_s
    =
    -\frac{2e^2}{\hbar^2}\bm\nabla_s^{{\mathsf T}}G_s\bm\nabla_s
    +
    V(\bm  s).
    \label{Hs}
\end{equation}

The lowest eigenstates of the effective Hamiltonian in Eq.~\ref{Hs} are obtained based on the determinant wavefunction ansatz~\cite{Pfau2024NES}. For $K_{\rm st}$ target states, we introduce $K_{\rm st}$ complex neural wavefunctions $\psi_j(\bm s)$ on the five-dimensional torus. A joint NES configuration is
\[
    \mathcal S =
    \left(\bm s^{(1)},\ldots,\bm s^{(K_{\rm st})}\right),
    \qquad
    \bm s^{(a)}\in[0,2\pi)^5 .
\]
The single-state wavefunctions are assembled into a matrix, and the total NES wavefunction is 
\begin{equation}
    {\cal M}_{aj}(\mathcal S)
    =
    \psi_j\!\left(\bm s^{(a)}\right),\qquad \Psi(\mathcal S)=\det{\cal M}(\mathcal S).
\end{equation}
The construction prevents the different variational states from collapsing onto the same state, without introducing explicit orthogonalization or overlap-penalty terms. After optimization, the state-resolved energies are obtained by diagonalizing the averaged energy matrix within the optimized low-energy subspace. 
Statistical uncertainties in the single-state energies are on the order of $0.01\,\text{GHz}$, and are estimated using a block jackknife over the final Monte Carlo energy-matrix estimates.

\subsection{Comparison between Harmonic Approximation and NN-VMC}

\begin{figure}[!ht]
    \centering
    \includegraphics[width=0.49\linewidth]{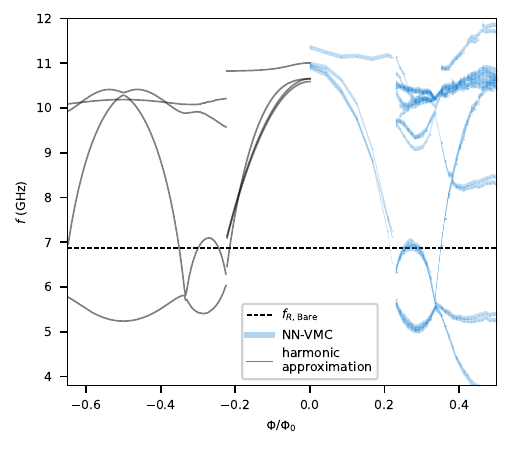}
    \caption{Comparison between harmonic approximation and NN-VMC calculation. Here we choose $E_J=63$ and $68\,\text{GHz}$ for the harmonic approximation and NN-VMC respectively, for each method to better align with experimental observations. }
    \label{fig:splitcomparison}
\end{figure}

In order to match the observed dome shape in the two-tone spectra (main text Fig.~\ref{fig:wide}c), 
as well as the four anticrossings between waffle and the readout resonator (main text Fig.~\ref{fig:s21}a), 
we use $E_J=63\,\text{GHz}$ for the harmonic-approximation calculation in Sec.~\ref{sec:harmonic}. 
The resulting spectrum is shown in Fig.~\ref{fig:splitcomparison}, and is compared side-by-side with the NN-VMC calculation of Sec.~\ref{sec:NNVMC}, 
which accounts for the anharmonicity of $V$ as well as tunneling between minima. 
A zoomed-in comparison near the CGS point is shown in main text Fig.~\ref{fig:wide}d. 
For the NN-VMC calculation a different $E_J$ value of $68\,\text{GHz}$ is used, 
which we obtain by matching the NN-VMC transition frequencies to the experimentally observed modes at $B=0\,\text{T}$. 
Compared to the harmonic approximation, NN-VMC is the more accurate numerical approach and shows better consistency with experiment, 
so $68\,\text{GHz}$ is a faithful estimate of the experimental $E_J$. 

We find that the harmonic approximation works well from zero flux up to the CGS point ($|\Phi/\Phi_0|\leq1/3$), 
generating eigenmodes that qualitatively follow the same trend as the NN-VMC results and the experiment. 
This suggests that inter-well tunneling is weak (tunneling rates much smaller than mode frequencies) in the four-minima and six-minima (CGS) regimes, 
as a result of the high potential barriers separating neighboring minima. 
However, tunneling becomes more significant in the two-minima regime, 
due to the decreasing potential barrier between the two global minima as $\Phi$ approaches $\Phi_0/2$. 
In the NN-VMC spectrum, the effect of tunneling between these two wells, specifically between their excited states, appears as the flat bands at roughly $5.5\,\text{GHz}$ and $8.5\,\text{GHz}$ near $\Phi_0/2$ (main text Fig.~\ref{fig:wide}b). 
These flat bands are observed in experiment (main text Fig.~\ref{fig:wide}a), but are missed by the harmonic approximation (Fig.~\ref{fig:splitcomparison}). 
NN-VMC is also able to reproduce the high-frequency band located above $11\,\text{GHz}$ near $\Phi_0/2$, consistent with experiment (main text Fig.~\ref{fig:wide}a,b).

\section{Origin of the fine structure in the two-tone spectrum}
\label{sec:finestructure}
This section justifies why two‑tone spectroscopy reveals the eigenmodes of all global minima, 
which may be degenerate at a given flux, and why local (shallower) minima can be neglected.
\subsection{Tunneling Between Minima}
In the multiple-minima regime, the tunneling rate between the ground states of different minima (estimated using the WKB approximation) is at least $10\,\text{kHz}$ -- 
smaller than the eigenfrequencies $\omega_{1\rightarrow5}$, which are on the order of $1\,\text{GHz}$, 
but still much larger than our $1\,\text{Hz}$ sampling rate in two-tone spectroscopy. 
This allows the waffle phases to hop many times among the degenerate minima during the collection of each two-tone data point. 
We note that the NN-VMC approach gives a tunneling rate on the order of $1\,\text{MHz}$ between ground states, 
but this value may be inaccurate since it is smaller than the typical numerical error of NN-VMC.
\subsection{Thermal Occupation}
Assuming that tunneling is sufficiently rapid to establish thermal equilibrium on the measurement timescale, the state of the waffle can be described by an incoherent mixture of the ground states of all minima:
$$\hat{\rho} = \sum_{\#} \frac{e^{-V_{\#}/k_{\!B}T}}{Q}\, |{\#}\rangle\langle{\#}|,
\qquad Q = \sum_{\#} e^{-V_{\#}/k_{\!B}T},$$
where $|\#\rangle$ denotes the ground state localized in minimum $\#$ and $V_{\#}$ is the potential energy at the bottom of that minimum. 
We note that at certain flux values close to the critical flux, such as $\Phi=-0.331$ and $\Phi=-0.335$, $V$ possesses global minima together with coexisting local minima that are higher in energy. 
However, according to the Boltzmann distribution above, only minima whose depth differs from that of the global minimum by less than $k_{\!B}T\approx1\,\text{GHz}$ carry significant weight. 
In our case, due to the large $E_J\approx68\,\text{GHz}$ of each Josephson junction, as well as the rapidly changing landscape of $V$ as a function of $\Phi$, the occupation of shallower local minima is negligible. 

\subsection{Evolution of the Potential Landscape with Flux}
\label{sec: landscape}
\begin{figure}[!ht]
    \centering
    \includegraphics[width=1\linewidth]{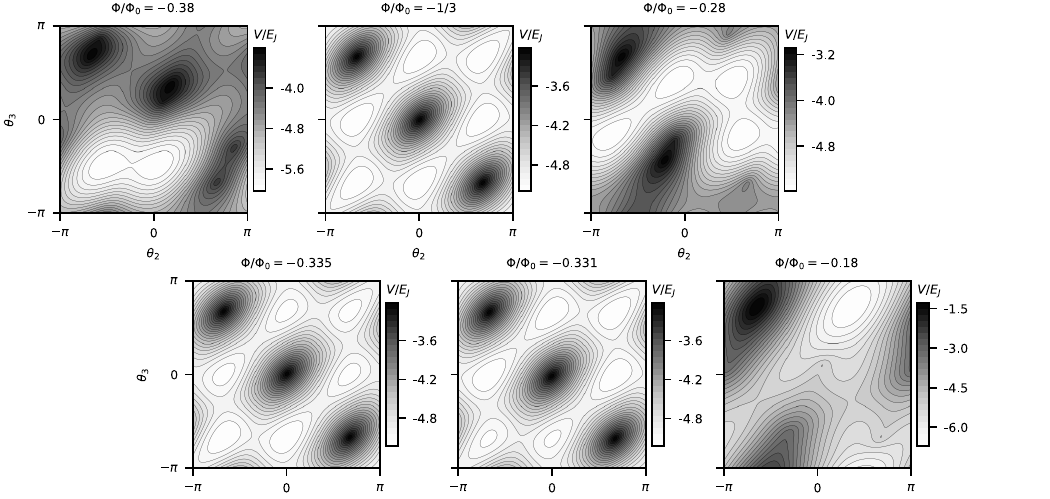}
    \caption{Josephson potential $V$ as a function of $\theta_2$ and $\theta_3$, plotted at representative flux values from $\Phi/\Phi_0=-0.18$ (right) to $\Phi/\Phi_0=-0.38$ (left).}
    \label{fig:potential}
\end{figure}

Figure~\ref{fig:potential} displays the classical potential $V(\theta_2,\theta_3)$ at several representative flux values. 
The number of global minima and the presence of local minima are summarized in Table~\ref{tab:minima}. 
The landscape changes rapidly near the critical point: a group of four wells and a group of two wells exchange stability, crossing exactly at $\Phi = -\Phi_0/3$, where all six become degenerate. 
The local minima that appear slightly away from the critical point are separated from the global minima by an energy larger than $k_{\!B}T$, so they are effectively unoccupied in our experiments.
\begin{table}[htbp]
    \centering
    \caption{Properties of the minima of \(V\) at representative flux values. Occupation probabilities are computed from the Boltzmann distribution at \(T = 50\,\text{mK}\).}
    \label{tab:minima}
    \begin{tabular}{|c|c|c|c|c|}
        \hline
        Flux \(\Phi\) (\(\Phi_0\)) & No. global minima & Occup. per global min. & No. local minima & Occup. per local min. \\
        \hline
        \(-0.18\)    & 1   & 1                 & --               & -- \\
        \hline
        \(-0.28\)    & 4   & 0.25 each         & --               & -- \\
        \hline
        \(-0.331\)   & 4   & 0.247 each        & 2                & 0.006 each \\
        \hline
        \(-1/3\) (CGS) & 6 & \(1/6 \approx 0.1667\) each & --          & -- \\
        \hline
        \(-0.335\)   & 2   & 0.48 each         & 4                & 0.01 each \\
        \hline
        \(-0.38\)    & 2   & 0.5 each          & --               & -- \\
        \hline
    \end{tabular}
\end{table}

\subsection{Consequences for the Two‑tone Spectrum}
Because the measurement averages over all thermally occupied global minima, every transition that is present in any of those minima can appear as a peak in the two‑tone response. 
If the capacitance and inductance matrices are perfectly symmetric, and if tunneling between different minima is neglected, the corresponding eigenfrequencies coincide and the peaks merge into a single line (as shown by the harmonic-approximation results in Fig.~\ref{fig:splitcomparison}). 
In a realistic device, tunneling between minima, coupling to the readout resonator, or slight disorder can break the exact degeneracy and cause each mode to split into a multiplet of closely spaced lines. 
The multiplicity of each two-tone band reflects the number of populated global minima. 

\subsection{Experimental Observation of Fine Structure}
\label{sec:finestructure_exp}
\begin{figure}[!ht]
    \centering
    \includegraphics[width=1\linewidth]{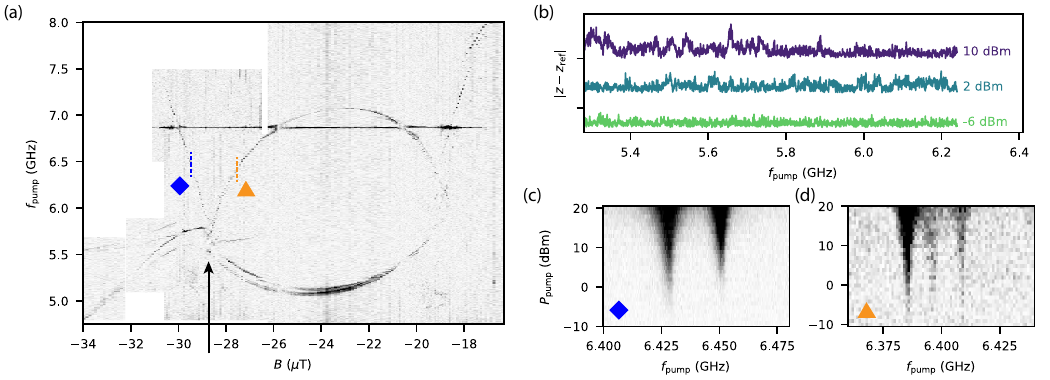}
    \caption{(a) Two-tone spectroscopy as a function of $B$, replotted from main text Fig.~\ref{fig:wide}(c).
    (b) Two‑tone spectrum at the critical CGS field $B=-28.7\,\mu\text{T}$ (indicated by the arrow in (a)), taken at various pump powers. The number of fine splittings is ambiguous.
    (c,d) Two-tone spectroscopy as a function of pump frequency and pump power, performed at the fields indicated by the dashed lines and symbols in (a). 
    In the low pump power limit, the mode splittings are threefold at the orange triangle and twofold at the blue diamond.}
    \label{fig:cgs_linecut}
\end{figure}

Beyond the mode splittings of main text Fig.~\ref{fig:power}, 
we examined additional fine structure at and near the CGS flux $\Phi = -\Phi_0/3$. 
At the CGS point the six minima are degenerate and the first and second excited states nearly coincide, 
so a 12-fold multiplet is expected. This exceeds our frequency resolution and signal-to-noise ratio, 
so we cannot determine the exact number of features at the CGS point (Fig.~\ref{fig:cgs_linecut}b). 

We also examined the higher-lying bands just to either side of CGS, 
where fewer minima contribute (Fig.~\ref{fig:cgs_linecut}c,d). 
Above CGS, where four minima are populated, 
both the harmonic approximation and NN-VMC predict a fourfold splitting, 
but we resolve only three lines (Fig.~\ref{fig:cgs_linecut}c). 
Below CGS, two minima are populated, so the harmonic approximation predicts a doublet, 
whereas NN-VMC predicts two additional modes from inter-well tunneling (four closely spaced modes, main text Fig.~\ref{fig:wide}d); 
experimentally we resolve a doublet (Fig.~\ref{fig:cgs_linecut}d).
In both cases the observed multiplicity in higher bands is lower than the prediction. 
The simplest explanation is that the missing lines are split by less than our frequency resolution. 
A second possibility is that some predicted transitions carry negligible spectral weight: 
if the matrix elements connecting the ground state to certain excited states are strongly suppressed, 
those lines would not appear as resolvable peaks, reducing the observed multiplicity. 

\section{Calculation of mode frequency of readout resonator coupled to waffle (harmonic approximation)}
\label{sec: readout}
Under the harmonic approximation, the total Lagrangian of a waffle coupled to a readout resonator is 
\begin{equation}
\label{eq: Lreadout}
\mathcal{L}_{W+R}=\mathcal{L}_\mathrm{HA}+\mathcal{L}_R+\mathcal{L}_\text{couple}=\mathcal{L}_\mathrm{HA}+\left(\frac{\hbar}{2e} \right)^2 \left(\frac{C_R\dot{\phi}_R^2}{2}-\frac{\phi_R^2}{2L_R}\right)+\left(\frac{\hbar}{2e} \right)^2 C_\text{couple}\,\dot{\phi}_R\,\dot{\phi}_1,
\end{equation}
where the waffle-only Lagrangian $\mathcal{L}_\mathrm{HA}$ is given by Eq.~\ref{eq: Lquadratic}, and $L_R$ and $C_R$ are the effective lumped-element inductance and capacitance of the resonator.
The relevant parameters for our readout resonator are 
resonant frequency $f_{R,\text{bare}}=6.87\,\text{GHz}$, characteristic impedance $Z_0=50\,\Omega$, $\left( \frac{\hbar}{2e} \right)^2 L_R^{-1}=222\,\text{GHz}$ and $C_R=728\,\text{fF}$.

We can diagonalize Eq.~\ref{eq: Lreadout} and find eigenmodes in a manner similar to the waffle-only case described in Sec.~\ref{sec:harmonic}, except that here we have one more degree of freedom (7 in total).
The $7\times7$ inverse inductance matrix $L^{-1}$ is simply the $6\times6$ version plus a seventh diagonal element equal to $1/L_R$. 
For the device shown in main text Fig.~\ref{fig:device}, the $7\times7$ capacitance matrix, including the capacitances between the waffle wires and the readout resonator, is obtained from the Ansys Q3D solver:
\begin{eqnarray}
C_{\text{ANSYS}}\,(\text{fF})=\begin{pmatrix}
  214.77 & -44.90 & -47.49 & -25.38 & -25.38 & -25.58 & -7.33 \\
  -44.90 & 191.28 & -44.98 & -25.39 & -27.00 & -25.43 & -0.34 \\
  -47.49 & -44.98 & 212.40 & -25.42 & -25.39 & -25.60 & -0.81 \\
  -25.38 & -25.39 & -25.42 & 212.18 & -50.49 & -48.40 & -0.23 \\
  -25.38 & -27.00 & -25.39 & -50.49 & 203.10 & -50.50 & -0.20 \\
  -25.58 & -25.43 & -25.60 & -48.40 & -50.50 & 211.97 & -0.14 \\
  -7.33  & -0.34  & -0.81  & -0.23  & -0.20  & -0.14  & 728 
\end{pmatrix}
\end{eqnarray}
Again, for the eigenmode calculation we assume ideal symmetry: 
(i) $C$ should be invariant under permutations of the matter wires, and 
(ii) the readout resonator couples only to gauge wire 1 of the waffle. 
We use the following symmetrized $C$ matrix:
\begin{eqnarray}
C\,(\text{fF})=\begin{pmatrix}
  214.77 & -44.90 & -47.49 & -25.5 & -25.5 & -25.5 & -7.5 \\
  -44.90 & 191.28 & -44.98 & -25.5 & -25.5 & -25.5 & 0\\
  -47.49 & -44.98 & 212.40 & -25.5 & -25.5 & -25.5 & 0\\
  -25.5 & -25.5 & -25.5 & 208 & -50 & -50 & 0\\
  -25.5 & -25.5 & -25.5 & -50 & 208 & -50 & 0\\
  -25.5 & -25.5 & -25.5 & -50 & -50 & 208 & 0\\
  -7.5  & 0 & 0 & 0 & 0 & 0 & 728
\end{pmatrix}
\end{eqnarray}
Following the same procedure as in Sec.~\ref{sec:harmonic}, we obtain the diagonalized Hamiltonian:
\begin{equation}
H_\mathrm{HA}=\sum_{i=1}^6\hbar\omega_i\left(\hat{a}_i^\dagger\hat{a}_i+\frac{1}{2}\right),
\end{equation}
where one of the modes $\omega_i$ is the resonator frequency dressed by the waffle in its ground state.

\begin{figure}[!p]
    \centering
    \includegraphics[width=0.49\linewidth]{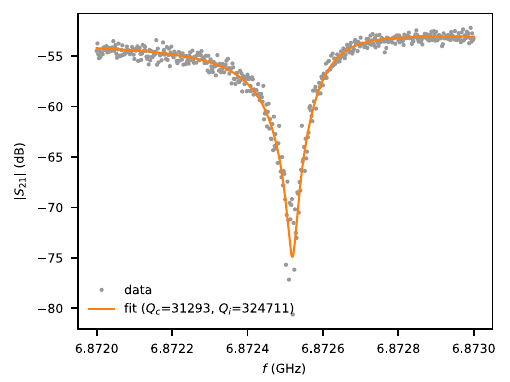}
    \caption{Resonator characterization at $B=0\,\text{T}$, where the waffle modes are far detuned. 
    At the signal power used here, there are approximately four photons in the resonator, which means it is well within the linear regime.
    From a circle fit we extract a coupling quality factor $Q_c\approx30\,000$ and an internal quality factor $Q_i\approx300\,000$.}
    \label{fig:res}
\end{figure}

\begin{figure}[!ht]
    \centering
    \includegraphics[width=0.80\linewidth]{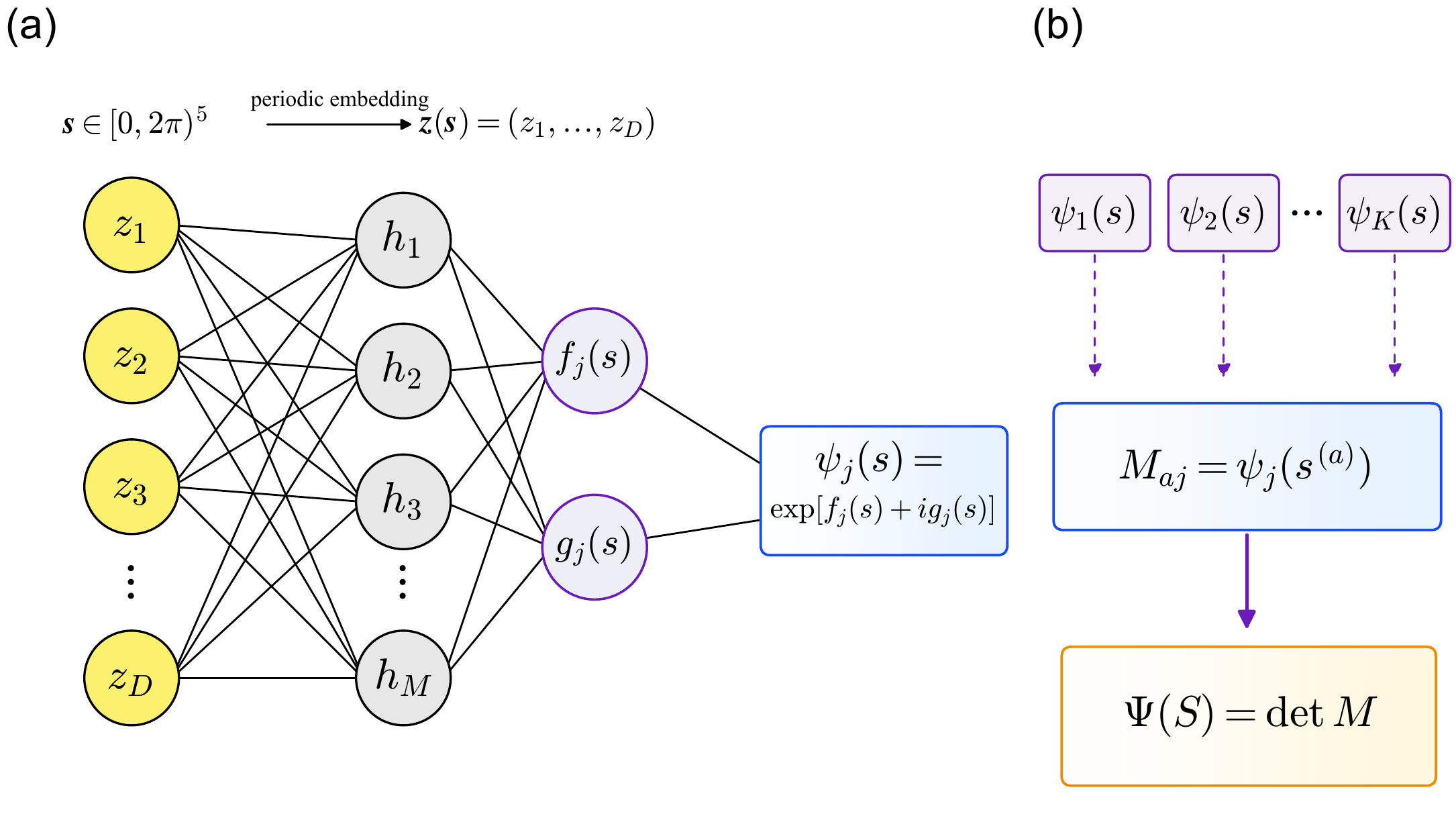}
    \caption{Architecture of the NN-VMC determinant wavefunction. 
    (a) A single component complex wavefunction $\psi_j(\bm{s})$. 
    The periodic coordinate $\bm{s}\in[0,2\pi)^5$ is first mapped to the embedded feature vector 
    $\bm{z}(\bm{s})=(z_1,\ldots,z_D)$, with input width $D=20$, and then passed through a one-hidden-layer neural network with hidden width $M=256$. 
    The two real outputs $f_j(\bm{s})$ and $g_j(\bm{s})$ define 
    $\psi_j(\bm{s})=\exp[f_j(\bm{s})+i g_j(\bm{s})]$. 
    (b) Construction of the NN-VMC determinant wavefunction from $K$ single-state components.}
    \label{fig:NN_VMC}
\end{figure}

\begin{figure}[!ht]
    \centering
    \includegraphics[width=1\linewidth]{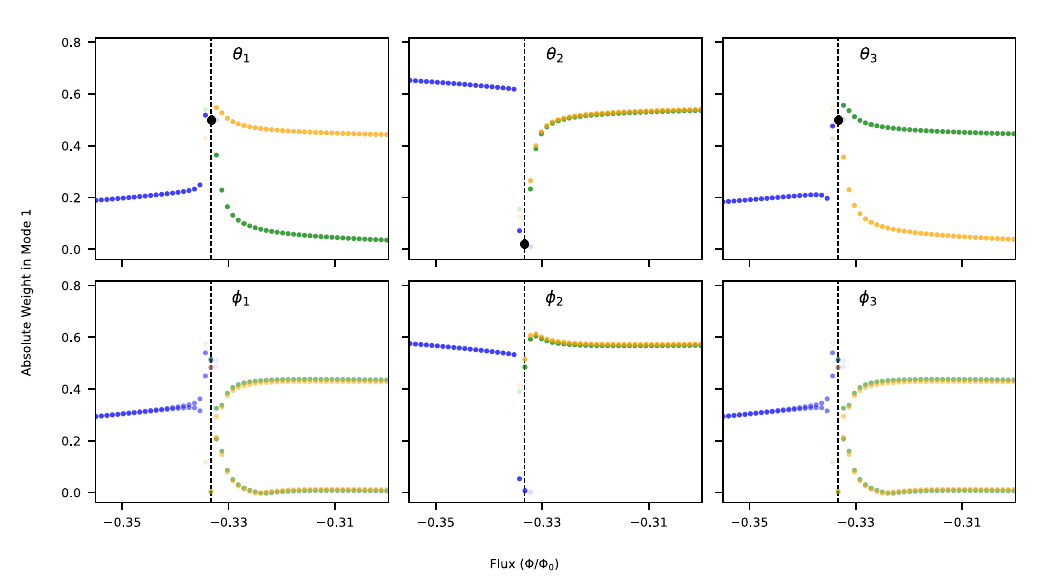}
    \caption{Calculated eigenvector of $\omega_1$ (the lowest mode) as a function of flux, using the harmonic approximation. 
    The top panels show the weights of the gauge wires $\theta_{1,2,3}$, while the bottom panels show the weights of the matter wires $\phi_{1,2,3}$. 
    Green/orange/blue colors correspond to the families of minima in main text Fig.~\ref{fig:s21}c. The dashed lines indicate the CGS point. 
    The eigenvectors of the six minima have the same gauge weights at the CGS point, represented by the black dots in the top row.}
    \label{fig:eigenvec}
\end{figure}

\begin{figure}[!ht]
    \centering
    \includegraphics[width=0.75\linewidth]{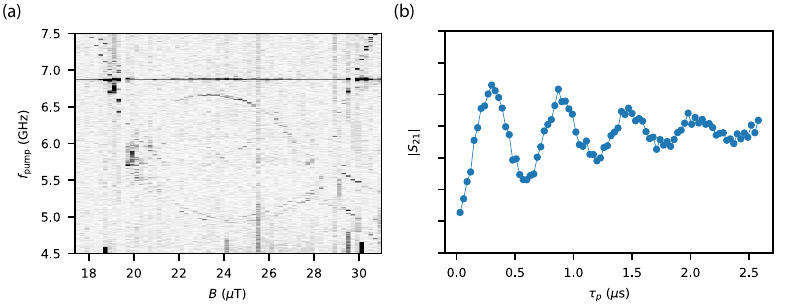}
    \caption{Measurements of a second waffle device of identical design. 
    (a) Two-tone spectroscopy as a function of magnetic field, showing a dome shape similar to that of the main device in main text Fig.~\ref{fig:wide}c.
    (b) Rabi oscillations as a function of the drive pulse duration $\tau_p$, taken at $B=+16\,\mu\text{T}$. 
    The drive frequency is set to $f_p=8045\,\text{MHz}$, equal to one of the waffle eigenmode frequencies.}
    \label{fig:dv2}
\end{figure}

\end{document}